\documentclass[aps,twocolumn,floats,pre]{revtex4}
\usepackage{graphics,graphicx,epsfig,color}
\usepackage{amssymb,amsmath}
\usepackage{epsf,wrapfig}
\usepackage[mathcal]{euscript}

\DeclareMathOperator{\Lagr}{\mathcal{L}}

\begin{document}

\title{Predictive information in a sensory population}

\author{Stephanie E. Palmer,$^a$
Olivier Marre,$^b$
Michael J. Berry II,$^b$ and William Bialek$^a$}

\affiliation{$^a$Joseph Henry Laboratories of Physics, $^a$Lewis--Sigler Institute for Integrative Genomics,  $^b$Department of Molecular Biology,  and $^b$Princeton Neuroscience Institute,
Princeton University, Princeton, New Jersey 08544}

\begin{abstract}
Guiding behavior requires the brain to make predictions about   future sensory inputs.  Here we show that efficient predictive computation starts at the earliest stages of the visual system. We estimate how much information groups of retinal ganglion cells carry about the future state of their visual inputs, and show that every cell we can observe participates in a  group of cells for which this predictive information is close to the physical limit set by the statistical structure of the inputs themselves.  Groups of cells in the retina also carry information about the future state of their own activity, and we show that this information can be compressed further and encoded by downstream predictor neurons, which then exhibit interesting feature selectivity.    Efficient representation of predictive information is a candidate principle that can be applied at each stage of neural computation.
\end{abstract}

\date{\today}

\maketitle

\section{Introduction}

Almost all neural computations involve making predictions.  Whether we are trying to catch prey, avoid predators, or simply move through a complex environment, the data we collect through our senses can guide our actions only to the extent that we can extract from these data information about the future state of the world.    Although it is natural to focus on the prediction of rewards \cite{schultz+al_97}, prediction is a much  broader problem, ranging from the seemingly simple extrapolation of the trajectories of moving objects to the learning of abstract rules that describe the unfolding pattern of events around us \cite{montague+sejnowski_94,rao+ballard_99,bnt}. An essential aspect of the problem in all these forms is that not all features of the past carry predictive power.    Since there are costs associated with representing and transmitting information, might sensory systems have developed coding strategies that are optimized, keeping only a limited number of bits of information about the past but ensuring that these bits are maximally informative about the future?  Could we go further, and imagine successive stages of signal processing by the brain as attempts to predict future patterns of neural activity?  Here we address these questions in the context of the vertebrate retina, taking advantage of new electrode arrays which make it possible to record simultaneously the activity of almost all the ganglion cells in a densely interconnected patch of the salamander retina \cite{marre+al_12}, giving us a nearly complete view through the brain's window on a small piece of the visual world.

\section{Coding for the position of a single visual object}

\begin{figure}[bth]
\centerline{\includegraphics[width=\linewidth]{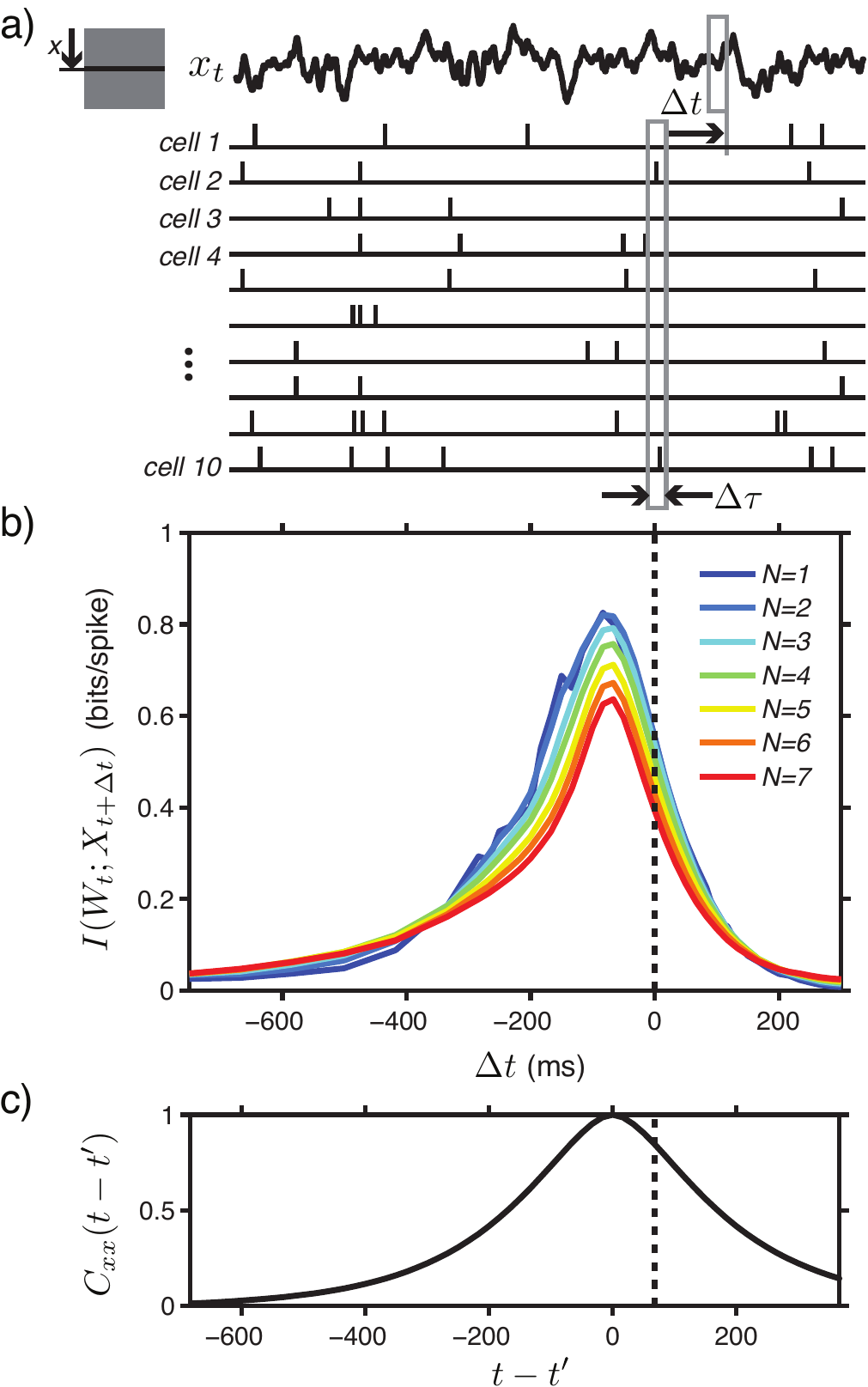}}
\caption{Information about past and future positions of a moving bar.   a) A schematic of the experiment.  Spiking responses are recorded from several cells simultaneously, and responses in a single small window of time $\Delta \tau$ can be expressed as binary words $w_t$.  These words provide information about the position of a moving bar at time $t' = t+ \Delta t$ into the past or future.  b)  Information that $N$--cell words about bar position [Eq (\ref{info1})],  as a function of the delay $\Delta t$.  Information is normalized by the mean number of spikes generated by each group of cells, and this is averaged over many $N$--cell groups to yield information in units of bits per spike in the population.   Standard errors of the mean over groups are negligible; estimates of information for a single group of cells have errors bars less than $0.01$ bits/spike. c) Autocorrelation function of the bar stimulus, $C_{xx}$, versus time delay, $t - t'$.  The zero-lag peak has been aligned with the peak information in part b). \label{x_info}}
\end{figure}

The structure of the prediction problem depends on the structure of the world around us.  In a world of completely random stimuli, for example, prediction is impossible.  To start, let us consider a relatively simple visual world such that, in the small patch of space represented by the neurons from which we record, there is just one object (a dark horizontal bar against a light background), moving along a trajectory $x_t$.   We want to imagine a world in which trajectories are predictable, but not completely; The moving object has some inertia, so that the velocities $v_t$ are correlated across time,  but is also ``kicked'' by unseen random forces.   A mathematically tractable example of a stochastic process with these properties is shown in Fig \ref{x_info}a,  along with the responses recorded from a population of ganglion cells in the salamander retina [see Eq.\ (\ref{model}) in Methods].

If we look at neural responses in small windows of time, e.g.  $\Delta \tau =1/60 \,{\rm s}$, almost all ganglion cells generate either zero or one action potential.  Thus, the activity of a single neuron, labeled $\rm i$, can be represented by a binary variable $\sigma_{\rm i}(t) = 1$ when the cell spikes at time $t$ and $\sigma_{\rm i}(t) = 0$ when it is silent.  The activity of $N$ neurons then becomes a binary ``word'' $w_t \equiv \{\sigma_1(t) ,\, \sigma_2 (t),\, \cdots ,\, \sigma_N(t)\}$.  If we (or the brain) observe the pattern of activity $w_t$ at time $t$, how much do we know about the position of the moving object?  Neurons are responding to the presences of the object, and to its motion, but there is some latency in this response, so that $w_t$ will be maximally informative about  the position of the object at some time in the past, $x_{t' < t}$.  On the other hand, we know that the brain is capable of solving the prediction problem, and that these ganglion cells provide all of the visual data on which such predictions are based, so it must be true that $w_t$ also provides some information about $x_{t'>t}$.

We can make these ideas precise by estimating, in bits, the information that the words $w_t$ provide about the position of the object at time $t'$ \cite{shannon_48,cover_book,spikes_book,bialek_12}:
\begin{equation} \label{info1} 
I(W_t;X_{t'}) =  \sum_{w_t,x_{t'}}  P_W(w_t)  P(x_{t'}|w_t)  \log_2 \left( \frac{P(x_{t'}|w_t)}{P_X(x_{t'})}\right) ,
\end{equation}
where $P_W(w)$ describes the overall distribution of words generated by the neural population,  $P_X(x)$ describes the distribution of positions of the object at one moment in time, and $P(x_{t'}|w_t)$ is the probability of finding the object at position $x$ at time $t'$ given that we have observed the response $w_t$ at time $t$.  We note that $P_X(x)$ is known, because we generate the trajectories, and an experiment of $\sim 1\,{\rm h}$ is sufficient to provide good sampling of the other distributions for populations of up to $N=7$ neurons (for details see Methods).   Results  are shown in Figure \ref{x_info}b, where we put the information carried by different numbers of neurons on the same scale by normalizing to information per spike.

As expected, the retina is most informative about the position of the object $t-t' = t_{\rm lat} \sim 80\,{\rm ms}$ in the past.  At this point, the information carried by multiple retinal ganglion cells is, on average, redundant, so that the information per spike declines as we examine the responses of larger groups of neurons.  Although the details of the experiments are different, the observation of coding redundancy at $t_\textrm{lat}$ is consistent with many previous results \cite{reich+al_01,petersen+al_01,puchalla+al_05,narayanan+al_05,schneidman+al_06,shlens+al_06,chechik+al_06,osborne+al_08,shlens+al_09,tkacik+al_09,schneidman+al_11,soo+al_11,doi+al_12}.  But the information that neural responses carry about position extends far into the past, $t' \ll t- t_{\rm lat}$, and more importantly this information extends into the future, so that the neural response at time $t$ {\em predicts} the position of the object at times $t' > t$.  This broad window over which we can make predictions and retrodictions is consistent with the persistence of correlations in the stimulus, as it must be (Fig \ref{x_info}c).  As we extrapolate back in time, or make predictions, the redundancy of the responses decreases, and there are hints of a crossover to synergistic coding of predictions far in the future, where the information per spike increases slightly as we look at larger groups of neurons.

\section{Bounds on predictability}

Should we be impressed by the amount of predictive information that is encoded by the retina?  In our simple world of one moving object, might prediction be ``just'' extrapolation, so that no special mechanisms are required?   Alternatively, might it be that the neurons make predictions about something other than the precise position of the object, so that what we see in Fig \ref{x_info}b is only part of the story?  To answer these questions we need to understand the limits to predictability.

Even if we keep a perfect record of everything we have experienced until the present moment, we cannot make perfect predictions:   all of the things we observe are influenced by causal factors that we cannot observe, and from our point of view the time evolution of our sensory experience thus has some irreducible level of stochasticity.  Formally, we imagine that we are sitting at time $t_{\textrm{now}}$ and have been observing the world, so far, for a period of duration $T$.  If we refer to all our sensory stimuli as $s(t)$, then what we have access to is the past $X_{\rm past} \equiv s(t_{\textrm{now}}-T< t  \leq t_{\textrm{now}})$.  What we would like to know is the future, $X_{\rm future} \equiv s(t >t_{\textrm{now}})$. The statement that predictive power is limited is, quantitatively, the statement that the predictive information, $I_{\rm pred}(T) \equiv I (X_{\rm past}; X_{\rm future})$ is finite \cite{bnt}.    This is the number of bits that the past provides about the future, and it depends not on what our brain computes but on the structure of the world.

\begin{figure}
\centerline{\includegraphics[width=0.7\linewidth]{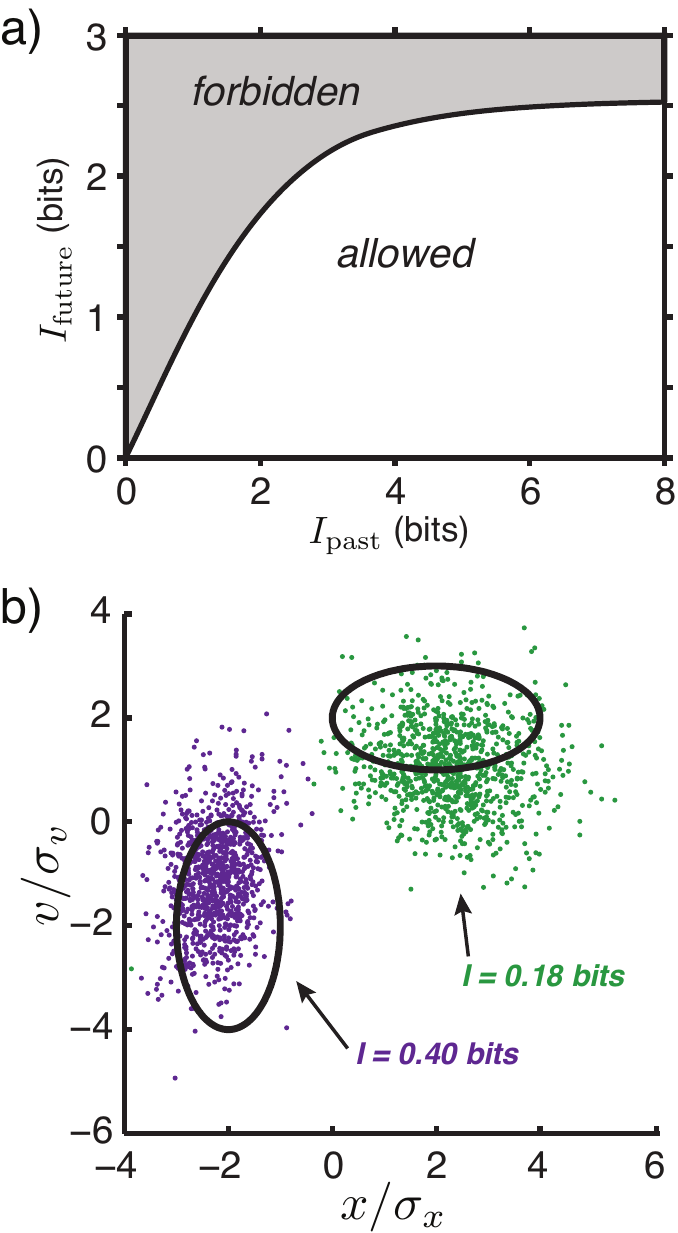}}
\caption{Bounds on predictive information. a)  Any prediction strategy defines a point in the plane $I_{\rm future}$ vs. $I_{\rm past}$.  This plane is separated into allowed and forbidden regions by a bound, $I_{\rm future}^*(I_{\rm past})$, which is shown for a sensory world of a single moving following the stochastic trajectories of Fog \ref{x_info}.  b) We can capture the same information about the past in different ways, as illustrated by the two black ``error ellipses'' in the position/velocity plane.  If we know that the trajectory is somewhere inside one of these ellipses, we have captured $I_{\rm past} = 0.42\,{\rm bits}$.  But points inside these ellipses propagate forward along different trajectories, and after $\Delta\tau = 1/60\,{\rm s}$ these trajectories arrive at the points shown in purple and green.  Using the same number of bits to make more accurate statements about position leads to more predictive  information ($I_{\rm future} = 0.40\,{\rm bits}$, purple) than if we use these bits to make more accurate statements about velocity ($I_{\rm future} = 0.18\,{\rm bits}$, green). \label{bound}}
\end{figure}

Not all aspects of our past experience are useful in making predictions.    Suppose that we build a compressed representation $Z$ of our past experience, keeping some features and throwing away others.  We can ask how much predictive information is captured by these features, $I_{\rm future} \equiv I(Z; X_{\rm future})$.  Notice that in building the representation $Z$ we start with our observations on the past, and so there is some mapping $X_{\rm past} \rightarrow Z$.  Further, this feature extraction captures a certain amount of information about the past, $I_{\rm past} \equiv I(Z; X_{\rm past})$.  The crucial point is that, given the statistical structure of our sensory world, $I_{\rm future}$ and $I_{\rm past}$ are related to one another.  Specifically, if we want to have a certain amount of predictive power $I_{\rm future}$, we need to capture a minimum number  of bits ($I^*$) about the past, $I_{\rm past} \geq I_{\rm past}^* (I_{\rm future})$.  Conversely, if we capture a limited number of bits about the past, there is a maximum amount of predictive power that we can achieve, $I_{\rm future} \leq I_{\rm future}^* (I_{\rm past})$, and we can saturate this bound only if we extract the most predictive features.  Thus,  we can plot information about the future vs. information about the past, and in any particular sensory environment this plane is divided into accessible and impossible regions; this is an example of the information bottleneck problem \cite{tishby+al_99,creutzig+al_09}.  In Fig \ref{bound}a we construct this bound for the simple sensory world of a ingle moving object, as in Fig \ref{x_info}.  To be optimally efficient at extracting information is to build a representation of the sensory world that is close to the bound which separates the allowed from the forbidden.

Building the maximally efficient predictor is nontrivial, even in seemingly simple cases.  For an object with trajectories as in Fig \ref{x_info}, knowledge of the object's position and velocity at time $t$ provides all the information possible about the future trajectory  (see Methods for details).  But knowing position and velocity exactly requires an infinite amount of information.  If, instead, we know the position and velocity only with some errors, we can draw an error ellipse in the position--velocity plane, as shown in Fig \ref{bound}b, and the area of this ellipse is related to the information that we have captured about the past.  Different points inside the error ellipse extrapolate forward to different trajectories, and if we look after a short time, an initial set of possibilities consistent with our limited information about the past has become a cloud of possible futures.  The key point is that error ellipses that have the same area but different shapes or orientations---using, for example, the limited number of available bits to provide relatively more or less information about positions vs.\ velocity---extrapolate forward to clouds of different sizes.  Thus, if we want to make the best predictions, we have to be sure that our budget of bits of about the past is used most effectively, and this is true even when prediction is ``just'' extrapolation.

\section{Direct measures of predictive information}

We would like to know if real neural populations reach the limits to predictability defined by Fig\ \ref{bound}: is the retina, in this sense, an efficient or even optimal encoder of predictive information?  Retinal ganglion cells encode both position and velocity of moving objects \cite{barlow_53,lettvin+al_60,maturana+al_60,barlow+al_64,roth_book,lee+al_93,frechette+al_05,thiel+al_07}, but it is not clear whether this encoding allows for optimal prediction.  To address this, we need to make a more general measurement of the predictive information carried by neural responses, not just the information about future positions as in Fig \ref{x_info}b.

\begin{figure}[b]
\centerline{\includegraphics[width=\linewidth]{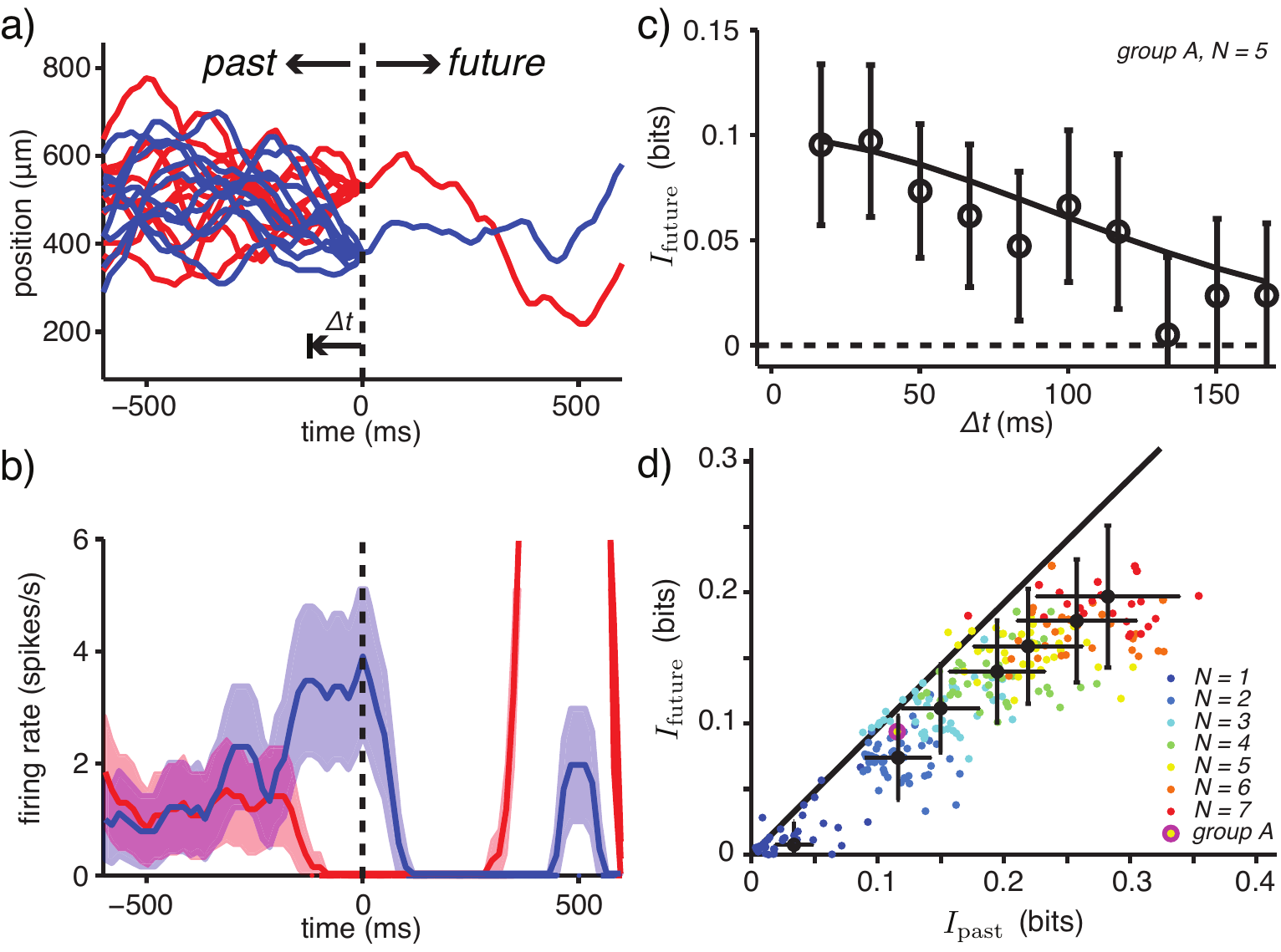}}
\caption{Direct measures of the predictive information in neural responses. a) Many independent samples of the trajectory $x_t$ converge onto one of several common futures, two of which are shown here (red and blue). The time of convergence is at $t = 0$, indicated by the vertical dashed line. b) Mean spike rates of a single neuron in response to the stimuli in a).  Shaded regions are $\pm$  one standard error in the mean.  are plotted as a blue or red shaded region.  For $t\ll 0$, responses are unmodulated and independent of the common future, while as we approach convergence at $t=0$ the cells respond to features that are specific to the future.  c) Information about the common future for one group of five cells, as a function of the time, $\Delta t$, until convergence.  Solid line shows the bound (see Methods).  d) Information about the future  vs. information about the past,  for many groups of different size, $N$;  group A is the 5-cell group from (c). Error bars and include contributions from the variance of the sample mean (across groups) and the standard deviation of the individual information estimates.  Solid line is the bound from Fig.\ \ref{bound}a.  
\label{commonF}}
\end{figure}

The statement that the neural response $w$ provides information about some feature $f$ of the stimulus means that there is a reproducible relationship between these two variables.  To probe this reproducibility we must present the same features many times, and sample the distribution of responses $P(w | f)$.   The information that $w$ provides about $f$ then is, on average \cite{shannon_48,cover_book,spikes_book,bialek_12},
\begin{equation}
I(W ; f) =  \sum_f P(f) \sum_{w} P(w|f) \log_2\left[ {{P(w|f)}\over{P_W (w)}}\right] ,
\label{IWf}
\end{equation}
where the features are drawn from the distribution $P(f)$ and the overall distribution of responses is given by
\begin{equation}
P_W(w) = \sum_f P(f) P(w | f) .
\end{equation}
In the case of interest here, the feature $f$ is the future of the stimulus.  To measure the information that neural responses carry about the future we thus need to repeat the future.  More precisely, we need to generate stimulus trajectories that are different, but converge onto the same future.  Given that we can write the distribution of trajectories $P[x(t)]$ (see Methods), we can draw multiple independent trajectories that have a ``common future,'' as shown schematically in Fig \ref{commonF}a \cite{bialek+al_06}.  In practice (see Methods for details), we synthesized one hundred independent pasts for each of thirty futures.

If trajectories converge onto a common future at time $t=0$, then for $t\ll 0$ the neural responses will be independent of the future, and we can see this in single cells as a probability of spiking that is independent of time or of the identity of the future (Fig \ref{commonF}b).  As we approach $t=0$, the neurons start to respond to aspects of the stimulus that are themselves predictive of the common future stimulus, and hence the probability of spiking becomes modulated.    Quantitatively, we can use Eq (\ref{IWf}) to estimate the information carried by responses from $N=1, 2, \cdots , 7$ neurons, as shown in Fig \ref{commonF}c for a particular 5-cell group.  We see that this predictive information is maximal when we try to make predictions over short times, and the predictive power gradually decays as we look farther into the future.

The particular group of five cells shown in Fig \ref{commonF}c captures $I_{\rm past} = 0.11\,{\rm  bits}$ of information about the past of the sensory stimulus, or $0.78\,{\rm bits/spike}$.  Figure \ref{bound} tells us that this amount of information is in the regime where the bound on prediction is almost linear with slope one---if the system is collecting the most useful bits, $0.11\,{\rm  bits}$ about the past can lead to $I_{\rm future}^*(I_{\rm past}) = 0.097 \,{\rm  bits}$ about the future.  In fact, this group of cells achieves a predictive $I_{\rm future}/I_{\rm future}^* = 0.98 \pm 0.39$, so that it is within error bars of being optimal.    We can generalize the bound in Fig \ref{bound} to ask what happens if we make predictions not of the entire future, but only starting $\Delta t$ ahead of the current time.  This prediction of more distant futures must be less reliable, and we can make this precise by computing the maximum predictive information as a function of $\Delta t$, again holding fixed the amount of information that is captured about the past.  We see that, for this one group of cells (Fig \ref{commonF}c), that the way in which predictive power decays as we try to extrapolate further into the future follows, within error bars, the theoretical limit set by the structure of the sensory inputs.

The results for the five--cell group in Fig \ref{commonF}c are not unusual.  For each of the 53 neurons in the population that we monitor, we can find a group of cells, including this neuron, that operates close to the bound in the $(I_{\rm past}, I_{\rm future})$ plane, as shown in Fig.\ \ref{commonF}d.  Not all groups that contain this neuron sit near the bound, but we do not expect a random sampling of cells to have this property.  How groups of cells are used downstream will determine which ones are polled as a group for predictive computations.  The fact that we have found that {\textit{every}} cell in this recording participated in {\textit{some}} group that sits near the bound is surprising. This continues to be true as we look at larger and larger groups of cells, until our finite data set no longer allows effective sampling of the relevant distributions.  At least under these stimulus conditions, populations of neurons in the retina thus provide near--optimal representations of predictive information, extracting from the visual input precisely those bits which allow maximal predictive power. 

\section{Predicting the future state of the retina}

When we monitor the activity of cells in the retina, it seems natural to phrase the problem of prediction in relation to the visual stimulus, as we have done in Fig \ref{commonF}.  But the brain has no access to visual stimuli except that provided by the retina.  For the central nervous system, then, predicting the future of visual inputs means predicting the future of retinal outputs.  

\begin{figure}
\includegraphics[width=\linewidth]{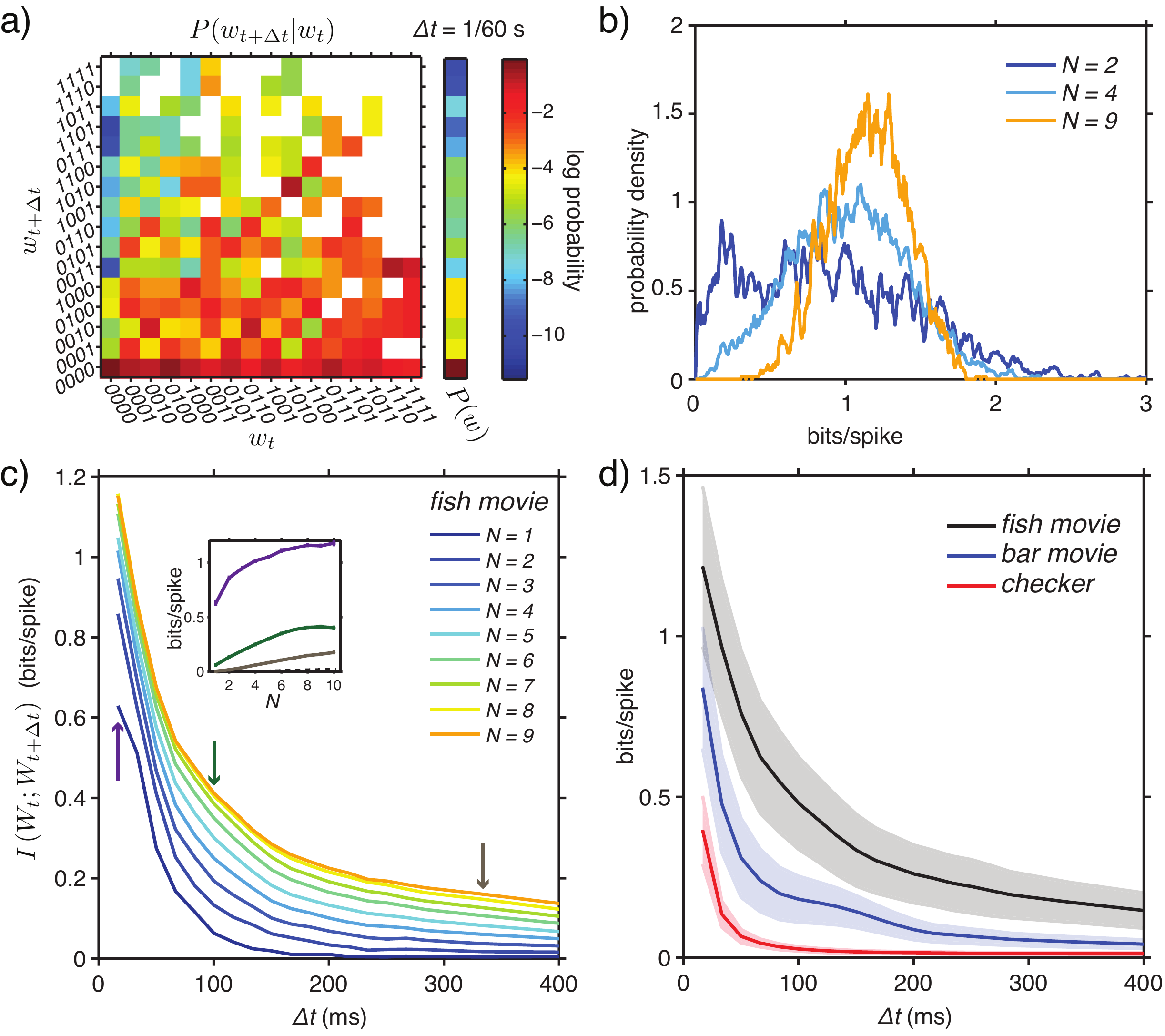}
\caption{Mutual information between past and future neural responses.  a) Conditional distribution of words, $P(w_{t+\Delta t}| w_t)$, at time $\Delta t = 1/60 \,{\rm s}$ for the group of 4 cells with the maximum information ($1.1\,{\rm bits/spike}$), in response to naturalistic movies.  The prior distribution of words, $P(w)$, is shown adjacent to the conditional.  Probabilities are plotted on a log scale; blank bins indicate zero samples. b) Distributions of $I (W_t;W_{t+\Delta t})$ for $N = 2$, $N= 4$, and $N=9$ cells, with $\Delta t = 1/60\,{\rm s}$.    c) Information between words  as a function of $\Delta t$, with $N$ ranging from 1 to 9.   Inset shown information vs. $N$ at $\Delta t$ marked by arrows. d) Information between words for groups of $N=9$, as a function $\Delta t$ for different classes of stimuli: natural movies (fish), the moving bar from Fig \ref{x_info}, and a randomly flickering checkerboard. Shaded regions indicate $\pm 1$ standard deviation across different groups of cells. \label{wordword}}
\end{figure}

If we observe that a population of retinal ganglion cells generates the word $w_t$  at time $t$, what can we say about the word that will be generated at time $t+\Delta t$ in the future?  We can give a complete, model independent answer to this question by sampling the conditional distribution of one word on the other, $P(w_{t+\Delta t}|w_t )$, and our ability to estimate this distribution is largely independent of the complexity of the visual inputs.  In Fig \ref{wordword}a we show an example  for $N=4$ cells, as the retina responds to naturalistic movies of underwater scenes (see Methods for details). The conditional probability of observing a particular word in the future given the observation of a word in the past is very different from the prior distribution of words (shown to the right), which means that there is significant mutual information between $w_t$ and $w_{t+\Delta t}$.  In Fig \ref{wordword}b, we show the distribution of this predictive information between words, for groups of $N=2$, $N=4$, and $N=9$ cells.  We have normalized the information in each group by the mean number of spikes, and we see that the typical bits/spike is growing  as we look at larger groups of cells.  Thus, the total predictive information in the patterns of activity generated by $N$ cells grows much more rapidly than linear in $N$: predictive information is encoded synergistically.

In Figure \ref{wordword}c we see how the predictive information varies both with the number of neurons and with the time delay over which the prediction is being made.  With these complex, naturalistic stimuli, larger groups of cells carry predictive information for hundreds of ms, and the maximum predictive information is well above $1\,{\rm bit/spike}$ on average across the thousands of groups that we sampled.   Importantly, smaller groups of cells do not carry long term predictive power, and even for short term predictions they carry roughly half the information per spike that we see in the larger groups.   

The large amounts of predictive information that we see in neural responses are tied to the structure of the sensory inputs, as shown in Fig \ref{wordword}d.   Naturalistic movies generate the most powerful, and most long ranged, predictions.  In contrast, the responses to random checkerboard movies lose predictability within a few frames, and motion of a single object (as in Fig \ref{x_info}) gives intermediate results.  The internal dynamics of the retina could generate predictable patterns of activity even in the absence of predictable structure in the visual world, but this doesn't seem to happen.  This raises the possibility that trying to predict the future state of the retina form its current state can lead us (or the brain) to focus on patterns of acidity that are especially informative about the visual world, and we will see that this is the case.

\section{Predictor neurons?}

The predictive information carried by $N$ neurons is more than $N$ times the information carried by single neurons, but even at $N=9$ it is less than one bit in total.   Can a neuron receiving many such ganglion cell inputs compress the description of the state of the retina at time $t$, while preserving the information that this state carries about what will happen at time $t+\Delta t$ in the future?  That is, can we do for the retinal output what the retina itself does for the visual input?  In particular,  if we can write down all of the predictive information in one bit, then we can imagine that there is a neuron inside the brain that takes the $N$ cells as inputs, and then a spike or silence at the output  of this `predictor neuron'  ($\sigma^{\rm out}$) captures the available predictive information.

\begin{figure*}
\includegraphics[width=0.9\linewidth]{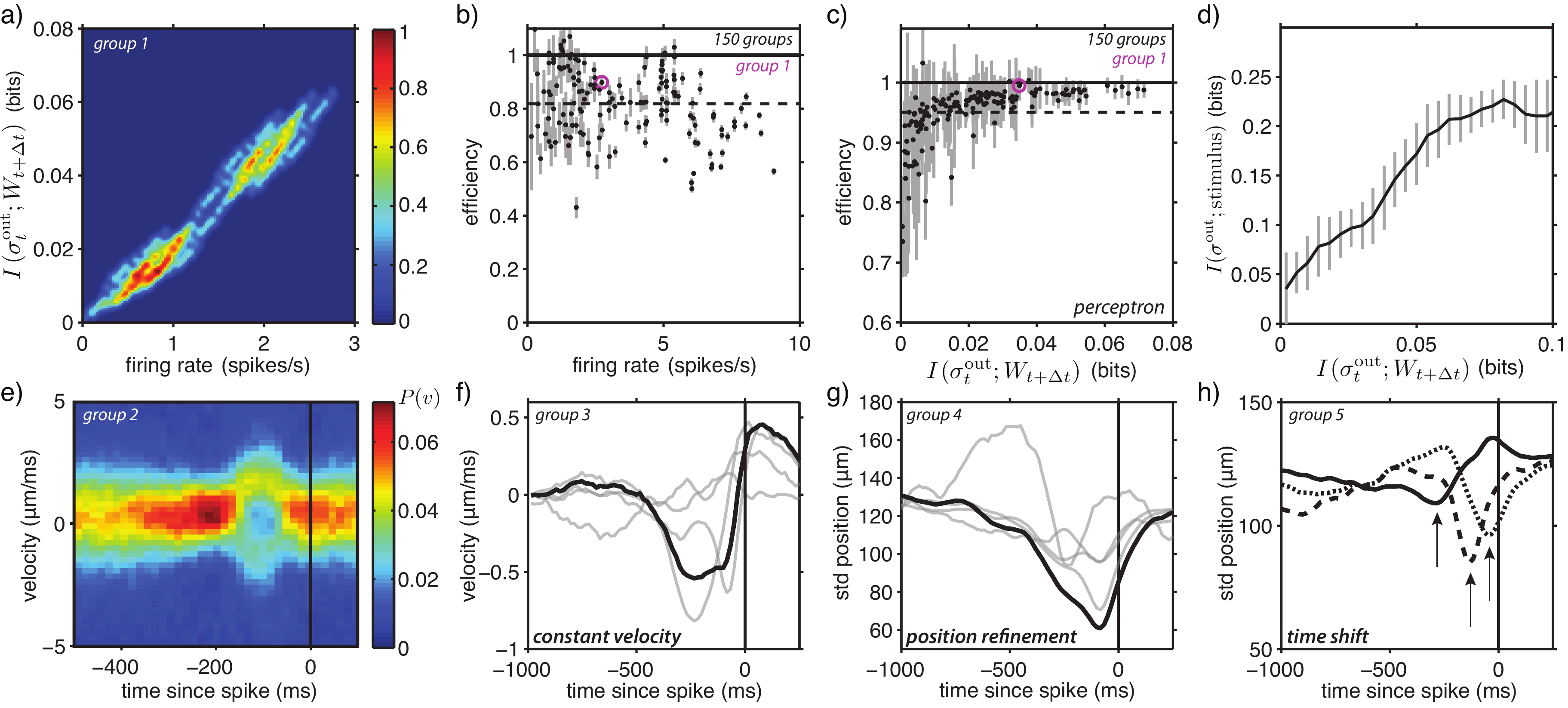}
\caption{Predictor neurons.   a) Predictive information, $I(\sigma^{\textrm{out}}_t;W_{t+\Delta t})$, captured by all possible mappings $w_t \rightarrow \sigma^{\rm out}$, as a function of the average firing rate of $\sigma^{\textrm{out}}$, for one particular four--cell input group.  Results are summarized as the local density of the 65,536 points in the plane, normalized to have a peak of 1.  b) The  maximum efficiency  $I(\sigma^{\textrm{out}}_t;W_{t+\Delta t})/I(W_t; W_{t+\Delta t})$ as a function of the output firing rate,  for 150 four--cell groups. Average over all groups is indicated by the dashed line, $y = .82$.  The solid black line, $y=1$, indicates perfect capture of all the input predictive information.   c) Efficiency of a perceptron  rule relative to the best possible rule, for the same groups as in (b).     d) The information that $\sigma^{\rm out}$ proves about  the visual stimulus grows with the predictive information that it captures.  Results are shown the means over  all possible output rules, for 150 four--cell input groups; error bars indicate standard deviations across the groups.    e) Distribution of stimuli that give rise to a spike in an optimized predictor neuron, for one particular group of four cells in response to the moving bar stimulus ensemble in Fig \ref{x_info}.  f) For a different four--cell group,  the average velocity triggered on a spike of the predictor neuron; light grey lines show the triggered averages for the input spikes; the predictor neuron selects for a long epoch of constant velocity.   g) For a third group of inputs,  the standard deviation of bar positions triggered on a spike in the predictor neuron (black) or on spikes in the individual input neurons (grey).   h) For a fourth group of cells, the standard deviation of bar positions conditional on a predictor neuron spike varies as we optimize for predictions with delays of $\Delta t = 1/30\,{\rm s}$ (solid curve), $\Delta t = 1/15 \,{\rm s}$ (dashed curve), and $\Delta t = 1/10\,{\rm  s}$ (dotted curve).
\label{sigout}}
\end{figure*}

Compressing our description of input words down to one bit means sorting the words $w_t$ into two groups $w_t \rightarrow \sigma^{\rm out}$, such that membership in the group is as predictive as possible about the word $w_{t+\Delta t}$.  If  this grouping is deterministic, then with $N$ neurons there are $2^{2^N}$ possible groupings;  with $N=4$, this is a manageable number ($65,536$), and so we can simply test all the possibilities, as shown in Fig \ref{sigout}a.  We see that it indeed is possible to represent almost all the predictive information from four neurons in the spiking or silence of a single neuron, and doing this does not require the `predictor neuron' to generate spikes at anomalously high rates.   This result generalizes across many groups of cells (Fig \ref{sigout}b), and we also find that the optimal rules can be well approximated by the predictor neuron thresholding a weighted sum of its inputs---a perceptron (Fig \ref{sigout}c). These results suggest that such predictor neurons are not only possible in principle, but biologically realizable.

Having extracted the predictive information from $N$ neurons,  we ask what this information means.  We emphasize that the predictor neurons are constructed without reference to the stimulus---just as the brain would have to do.  By providing an unsupervised solution to the prediction problem, have we made progress toward understandable computations on the visual inputs?  By repeating the same naturalistic movie many times, we can measure the information that the spiking of a predictor neuron carries about the visual input, using standard methods \cite{brenner+al_00,schneidman+al_11}.  As we see in Fig \ref{sigout}d, model neurons that extract more predictive information also provide more information about the visual inputs.  Thus, by solving the prediction problem, the brain can ``calibrate'' the combinations of spiking and silence in the ganglion cell population, grouping them in ways that capture more information about the visual stimulus.

What features of the visual input are being represented when we extract the predictive information?  To answer this question we return to the simple world of a single bar moving on the screen, as described above.  Now we can ask for the distribution of trajectories $x_t$ conditional on a spike at the output of the predictor neurons, and examples of this are shown in Fig \ref{sigout}e through h.  We see that the predictor neurons extract aspects of stimulus motion---motion at constant speed but either direction (Fig \ref{sigout}e) and long epochs of constant speed (Fig \ref{sigout}g). The estimate of the bar position in the predictor neurons is better (lower variance) than in any one of its inputs (Fig \ref{sigout}f), showing that optimizing for predicting inputs leads to a refinement in the stimulus estimate.  Also, these downstream cells have interesting feature selectivity when they are optimized (for the same inputs) to make predictions farther into the future (Fig \ref{sigout}h):  the time of sharpest stimulus discrimination moves closer to the time of a spike in the downstream cell when it is more predictive of its inputs farther in the future, so that searching for predictable features can lead to compensation of latencies.  Thus, searching for efficient representations of the predictive information in the state of the retina itself drives the emergence of motion estimation.  This supports the intuition that the visual system computes motion not for its own sake, but because, in a world with inertia, motion estimation provides an efficient way of representing the future state of the world.

\section{Discussion}

The classical approach to the analysis of sensory coding and information processing focuses on those features of the sensory input which trigger action potentials in neurons at various levels of the brain's processing hierarchy.  While much has been learned in this way, there are two important aspects of the coding problem as `seen' by the organism itself that are missed.  First, to correlate spikes with sensory stimuli we need independent access to the stimulus, which the brain does not have.  Second, because sensory systems are causal, the stimulus features which trigger spikes are events that occurred in the past, but to guide its actions the organism needs to make estimates about what will happen in the future.  By focusing on the representation of predictive information, we can address both of these issues.

Since Shannon's foundational work, it has been hoped that information theory would provide not just a guide to the design of manmade communication systems, but also a framework for understanding the representation and processing of information in naturally occurring systems, including the brain.  A central problem is that while one can speak of the entropy in a signal, without reference to the meaning or value of those bits, ``information'' is always information about {\em something}.  One might argue that the successful applications of information theoretic ideas to biological systems are in those cases where it is clear which information is relevant.  But how can we use information theoretic ideas more generally?  While it is difficult to guess how organisms will value information about particular features of the world, we know that value can be attached only to bits that have the power to predict the organism's future sensory experience.  Importantly, these predictive bits are a tiny fraction of the total number of bits that our sensory systems can collect, and so simply providing an efficient representation of the predictive information---separating the potentially valuable from the surely valueless---may take the system a long way toward its goal.  By estimating how much information neural responses provide about the future of sensory stimuli, even in a simple world, we have found evidence that the retina really does provide an efficient, and perhaps nearly optimal, representation of predictive information.  This optimization principle is very different from classical ideas about the reduction of redundancy or the maximization of the (total) information transmission, and it seems that we can distinguish among these candidate principles experimentally.

The efficient representation of predictive information is a principle that can be applied not just to retinal coding of visual inputs, but at every layer of neural processing.  As an illustration, we consider the problem of a single neuron that tries to predict the future of its inputs from other neurons, and encodes its prediction in a single output bit---spiking or silence.  This problem provides a way of analyzing the responses from a population of neurons that makes no reference to anything but the responses themselves, and in this sense provides a model for the kinds of computations that the brain can do.  The maximally efficient representations that we find involve processing which generates output spikes at a reasonable rate, even without any further constraints, and the structure of these computations is simple enough to be learnable by biologically plausible rules.  The optimal `predictor neurons' also are efficient transmitters  of information about the sensory input, even though the rules for optimal prediction are found without looking at the stimulus.  Thus, solving the prediction problem allows the central nervous system to identify features of the retina's combinatorial code that are especially informative about the visual world, without any external calibration.  Finally, optimizing the representation of predictive information drives the emergence of neurons selective for features of the visual world, such as motion, which are known to be relevant at various stages along the visual pathway.  

The idea that neural coding of sensory information might be efficient, or even optimal, in some information theoretic sense, is not new.  Individual neurons have a capacity to convey information that depends on the time resolution with which spikes are observed, and one idea is that this capacity should be used efficiently \cite{mackay+mcculloch_52,rieke+al_93}.  Another idea is that the neighboring cells in the retina should not waste their capacity by transmitting redundant signals, and minimizing this redundancy may drive the emergence of spatially differentiating receptive fields \cite{barlow_61,attneave_54,atick+redlich_92}.   In a similar vein, temporal filtering may serve to minimize redundancy in time \cite{dan+al_96}, and this is sometimes called ``predictive coding'' \cite{srinivasan+al_82,  berry+schwartz_11}.  In the simplest implementations of predictive coding, one identifies the predictable components of the signal and removes these, encoding only the deviations from expectation, or ``surprises.''  In contrast, having immediate access to predictive information requires the opposite, an encoding of those features of the past which provide the basis for optimal prediction.  The efficient coding of predictive information thus seems to be a very different principle from those articulated previously, and rests on the assignment of value to information about the future. 

While there has been much interest in the brain's ability to predict particular things---rewards \cite{schultz+al_97}, the reversal of motion \cite{berry+al_99}, the next occurrence of a periodic stimulus \cite{schwartz+al_07}---our approach emphasizes that prediction is a general problem, which can be stated in a unified mathematical structure across many different contexts.  Once we know the context, as defined by the statistical structure of sensory inputs, including inputs about reward and the success or failure of different actions, the general problem takes on a more specific structure.  But prediction includes everything from the extrapolation of trajectories to the learning of rules, so that finding maximally efficient representations of the predictive information corresponds to finding optimal solutions to these diverse, biologically relevant, problems \cite{bialek+al_06}.  Our results on the efficient representation of predictive information in the retina thus may hint at a much more general principle.

\begin{acknowledgments}
We thank E Schneidman, GJ Stephens and G Tka\v{c}ik for useful discussions, and GW Schwartz, D Amodei, and FS Soo for help with the experiments.    The work was supported in by National Science Foundation Grants IIS--0613435, PHY--0957573, and CCF--0939370, by  National Institutes of Health  Grant EY--014196, by Novartis (through the Life Sciences Research Foundation), by the Swartz Foundation, and by the WM Keck Foundation.  We also thank  the Aspen Center for Physics, supported by NSF grant PHY--1066293, for its hospitality.
\end{acknowledgments}

\bigskip\bigskip\hrule\bigskip

{\footnotesize\noindent Current addresses:  
\begin{itemize}
\item[SEP:] Department of Organismal Biology and Anatomy, The University of Chicago, Chicago, IL 60637.
\item[OM:]  Institut de la Vision, UMRS 968 UPMC, INSERM, CNRS U7210, CHNO
Quinze--Vingts, F--75012 Paris, France.
\end{itemize}}

\bigskip\hrule\bigskip

\section*{Methods}

{\em Mutli--electrode recordings.}
Data were recorded from larval tiger salamander retina using the dense  252--electrode arrays with $30\,\mu{\rm m}$ spacing, as described in \cite{marre+al_12}.  A piece of retina was freshly dissected and pressed onto the multi--electrode array. While the tissue was perfused with oxygenated artificial cerebral spinal fluid, images from a computer monitor were projected onto the photoreceptor layer via an objective lens.  Voltages were recorded from the 252 electrodes at $10\,{\rm kHz}$ throughout the experiments, which lasted 4 to 6 hours.  Spikes were sorted conservatively \cite{marre+al_12}, yielding populations of 49 or 53 identified cells from two experiments, from which groups of different sizes were drawn for analysis, as described in the text and below. 

{\em Stimulus generation and presentation.}  Movies were presented to the retina from  $360\times 600$ pixel display, with 8 bits of greyscale.  Frames were refreshed at $60\,{\rm fps}$ for naturalistic and moving bar stimuli, and at $30\,{\rm fps}$ for randomly flickering checkerboards.  The monitor pixels were square and had a size of $3.81\,\mu{\rm m}$ on the retina.
The moving bar (Fig \ref{x_info}) was 11 pixels wide and black (level 0 on the greyscale) against a background of grey (level 128).  The naturalistic movie was a $19\,{\rm s}$ clip  of fish swimming in a tank during feeding on an algae pellet, with swaying plants in the background, and was repeated a total of 102 times. All movies were normalized to the same mean light intensity.

{\em Motion trajectories.}  The moving bar stimulus was generated by a stochastic process that is equivalent to the Brownian motion of a particle bound by a spring to the center of the display:  the position and velocity of the bar at each time $t$ were updated according to
\begin{eqnarray}  \label{model}
x_{t+\Delta \tau} &=& x_t +  v_t \Delta\tau \label{x_dyn} \\ \nonumber
v_{t+\Delta\tau} &=& [1- \Gamma\Delta\tau] v_t - \omega^2 x_t\Delta\tau  +  \xi_t \sqrt{D\Delta\tau} ,  
\end{eqnarray}
where $\xi_t$ is a Gaussian random variable with zero mean and unit variance, chosen independently at each time step.   The natural frequency $\omega = 2\pi \times (1.5 \,{\rm s}^{-1})\textrm{rad}/\textrm{s})$, and the damping $\Gamma = 20 \,{\rm s}^{-1}$; with $\zeta = \Gamma/2\omega = 1.06$, the dynamics are slightly overdamped.  The time step $\Delta\tau =1/60\,{\rm s}$ matches the refresh time of the display, and we chose $D = 1\,{\rm pixel}^2/{\rm s}^3$ to generate a reasonable dynamic range of positions.  Positions at each time were rounded to integer values, and we checked that this discretization had no significant effect on any of the statistical properties of the sequence, including the predictive information.

{\em Common futures.}  To create  trajectories in which several independent pasts converge onto a  common future, we first generated a single very long trajectory, comprised of 10 million time steps. 
From this long trajectory we searched for segments with a length of 52 time steps such that the last two positions in the segment were common across multiple segments, and we joined each of these ``pasts'' on to the same future, generated with the common endpoints as initial conditions.  30 such distinct futures with their associated 100 pasts were identified for display during experiments.  Both the past and the future segments of the movie were each $50\Delta\tau \sim 833\,{\rm ms}$. in duration.  Past--future clips were presented in pseudorandom order.

{\em Estimating information.}  For all mutual information measures, we followed Ref \cite{strong+al_98}:  data were subsampled via a bootstrap technique for different fractions $f$ of the data,  with 50 bootstrap samples taken at each fraction.  For each sample we identify frequencies with probabilities, and plug in to the definition of mutual information to generate estimates $I_{\rm sample}(f)$. Plots of $I_{\rm sample}(f)$ vs. $1/f$ were extrapolated to infinite sample size ($1/f \rightarrow 0$), and the intercept $I_\infty$ is our estimate of the true information; errors were estimated as the standard deviation of $I_{\rm sample}(f)$ at $f=0.5$,  divided by $\sqrt{2}$.   Information estimates also were  made for randomly shuffled data, which should yield zero information.  If the information from shuffled data differed from zero by more than the estimated error, or by more than absolute cutoff of $0.02\,{\rm bits/spike}$, we concluded that we did not have sufficient data to generate a reliable estimate.  In estimating information about bar position (Fig \ref{x_info}), we compressed the description of position into $K=37$ equally populated bins, and checked that the information was on a plateau vs. $K$.  When we compute the information that neural responses carry about the past stimulus, we follow Refs \cite{brenner+al_00,schneidman+al_11}, making use of the repeated ``futures'' in the common future experiment. 

\begin{figure}[t]
\includegraphics[width=0.8\linewidth]{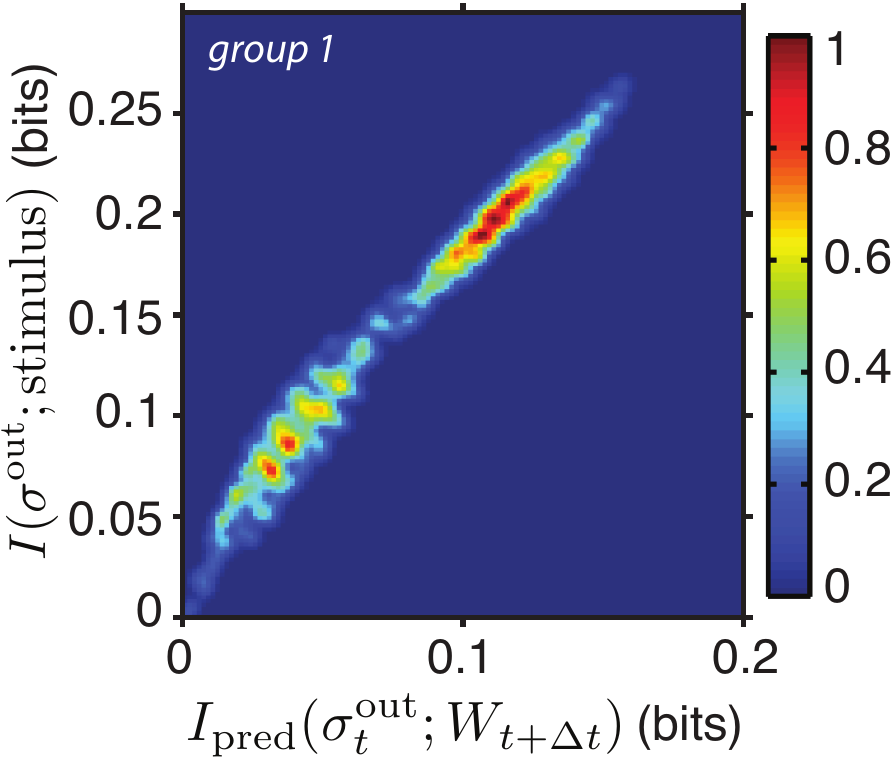}
\caption{Increasing predictive information enhances stimulus coding for single output cells.  The stimulus information for all downstream rules for one particular 4-cell input group are plotted as a density plot.  The scale bar on the right indicates density, normalized to a peak of 1.
\label{S1}}
\end{figure}

{\em Information bottleneck.} 
Information about the future of the stimulus is bounded by the optimal compression of the past, for each given compression amount.  Formally, we want to solve the ``bottleneck problem'' \cite{tishby+al_99}:
\begin{equation}
\min\limits_{p(z|x_{\rm past})}\Lagr = I(X_{\rm past};Z) - \beta I(Z;X_{\rm future}) ,
\label{bottleneck}
\end{equation}
where we map pasts $x_{\rm past} \in X_{\rm past}$ into some compressed representation $z\in Z$, using a probabilistic mapping $p(z|x_{\rm past})$.  The parameter $\beta$ sets the tradeoff between compression (reducing the information that we keep about the past, $I(X_{\rm past};Z)$) and prediction (increasing the information that we keep about the future, $I(Z;X_{\rm future})$). Once we find the optimal mapping, we can plot $I(Z;X_{\rm future})$ vs. $I(X_{\rm past};Z)$ for the one parameter family of optimal solutions obtained by varying $\beta$.  In general this is a hard problem.  Here we are interested in trajectories such that position and velocity (together) are both Gaussian and Markovian, from Eqs (\ref{model}).  The Markovian structure means that optimal predictions can always be based on information contained at the most recent point in the past, and that prediction of the entire future is equivalent to prediction one time step ahead.  Thus we can take $x_{\rm past} \equiv (x_t, v_t)$ and $x_{\rm future} \equiv (x_{t+\Delta\tau} , v_{t+\Delta \tau})$.    The fact that all the relevant distributions are Gaussian means that there is an analytic solution to the bottleneck problem \cite{chechik+al_05}, which we used here.


\section*{Supplementary Information}

{\em Stimulus information in $\sigma^\textrm{out}$ for one group.}
In Figure 5d, we plotted the average stimulus information as a function of predictive information about the future inputs for 200 downstream cells.  In Fig \ref{S1} we plot the same information for one group of 4 retinal input cells and all possible binary output rules.  This shows that capturing more of the predictive information in the patterns of retinal ganglion cell activity also allows the hypothetical predictor neuron to convey greater information about the visual stimulus: building better local predictions leads to better stimulus coding.

\begin{figure*}[t]
\includegraphics[width=0.9\linewidth]{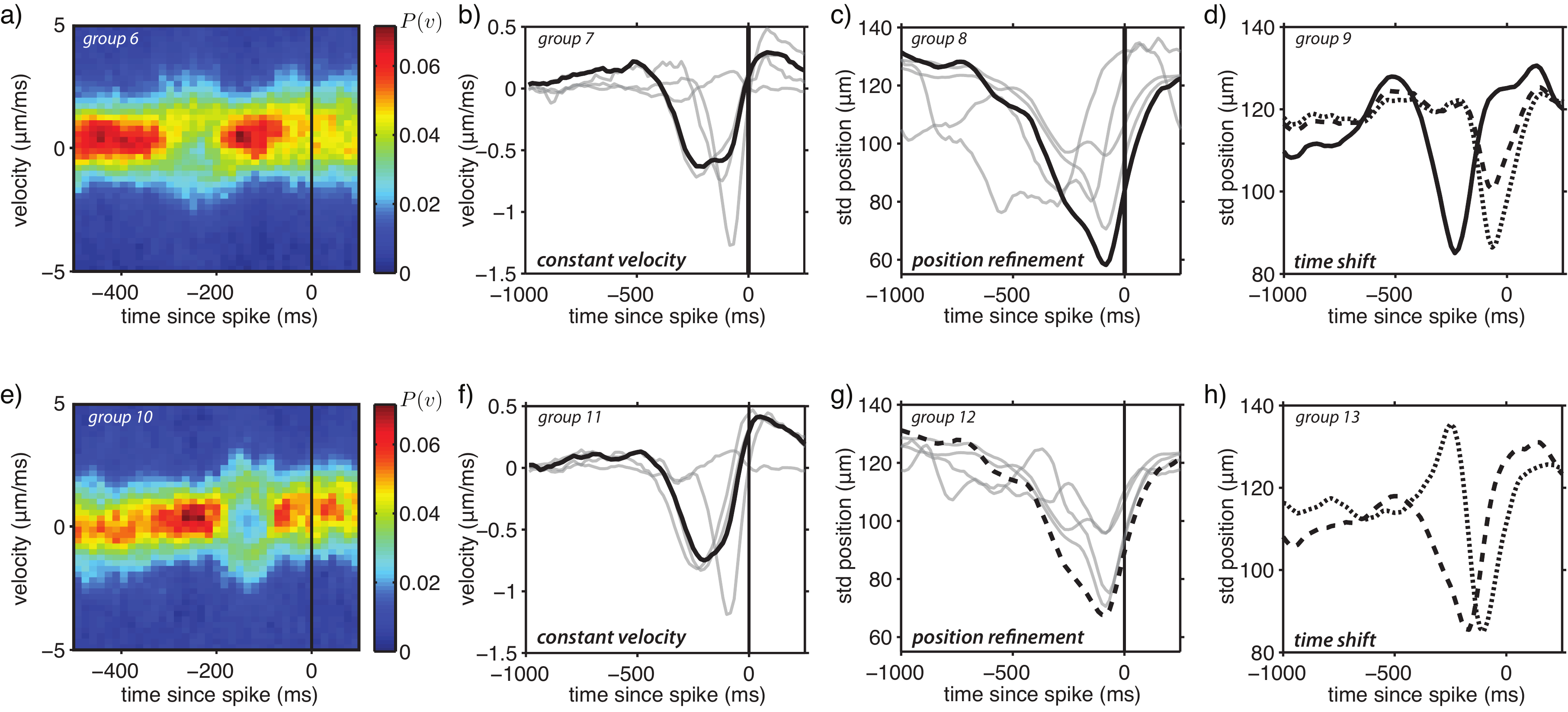}
\caption{Stimulus features extracted by firing in optimized predictor neurons: more examples on the theme of Fig \ref{sigout}. a) and e) Distribution of stimuli that give rise to a spike in an optimized predictor neuron, for two particular groups of four cells in response to the moving bar stimulus ensemble in Fig \ref{x_info}; the predictor neuron selects for motion at constant speed, with relatively little direction selectivity.  b) and f) The average velocity triggered on a spike of the predictor neuron; the predictor neurons select for a long epoch of constant velocity.   Light grey lines show the triggered averages for the input spikes.  c) and g) The standard deviation of bar positions triggered on a spike in the predictor neuron (black) or on spikes in the individual input neurons (grey); predictors neurons provide a more refined position estimate.   d) and h) The standard deviation of bar positions conditioned on a predictor neuron spike varies as we optimize for predictions with delays of $\Delta t = 1/30\,{\rm s}$ (solid curve), $\Delta t = 1/15 \,{\rm s}$ (dashed curve), and $\Delta t = 1/10\,{\rm  s}$ (dotted curve); optimizing predictions can compensate for latencies.
\label{S2}}
\end{figure*}

{\em Feature selectivity in predictor neurons.}
In Figure 5e-h, we showed four kinds of stimulus feature selectivity that emerged in our analysis of  optimized predictor neurons.  In Fig \ref{S2} we show two more examples for each of these interesting features.


\begin{thebibliography}{99}
\expandafter\ifx\csname natexlab\endcsname\relax\def\natexlab#1{#1}\fi
\expandafter\ifx\csname url\endcsname\relax
  \def\url#1{\texttt{#1}}\fi
\expandafter\ifx\csname urlprefix\endcsname\relax\def\urlprefix{URL }\fi

\bibitem{schultz+al_97}
Schultz, W., Dayan, P. \& Montague, P.~R.
\newblock A neural substrate of prediction and reward.
\newblock \emph{Science} \textbf{275}, 1593--9 (1997).

\bibitem{montague+sejnowski_94}
Montague, P.~R. \& Sejnowski, T.~J.
\newblock The predictive brain: temporal coincidence and temporal order in
  synaptic learning mechanisms.
\newblock \emph{Learning \& Memory} \textbf{1}, 1--33 (1994).

\bibitem{rao+ballard_99}
Rao, R.~P. \& Ballard, D.~H.
\newblock Predictive coding in the visual cortex: a functional interpretation
  of some extra-classical receptive-field effects.
\newblock \emph{Nature Neurosci} \textbf{2}, 79--87 (1999).


\bibitem{bnt}
Bialek, W., Nemenman, I. \& Tishby, N.
\newblock Predictability, complexity and learning.
\newblock \emph{Neural Comp} \textbf{13}, 2409--2463 (2001).

\bibitem{marre+al_12}
Marre, O. \emph{et~al.}
\newblock Mapping a complete neural population in the retina. 
\newblock \emph{J Neurosci} \textbf{32}, 14859--73 (2012).


\bibitem{shannon_48}
Shannon, C.~E.
\newblock A mathematical theory of communication.
\newblock \emph{Bell Sys Tech J} \textbf{27}, 379--423 \& 623--656 (1948).

\bibitem{cover_book}
Cover, T.~M. \& Thomas, J.~A.
\newblock \emph{Elements of Information Theory} (Wiley, New York, 1991).

\bibitem{spikes_book}
Rieke, F., Warland, D., de~Ruyter~van Steveninck, R.~R. \& Bialek, W.
\newblock \emph{Spikes:\! Exploring the Neural Code}.
\newblock  (MIT Press, Cambridge, 1997).

\bibitem{bialek_12}
Bialek, W.
\newblock \emph{Biophysics: Searching for Principles} (Princeton University
  Press, Princeton, 2012).

\bibitem{reich+al_01}
Reich, D.~S., Mechler, F. \& Victor, J.~D.
\newblock Independent and redundant information in nearby cortical neurons.
\newblock \emph{Science} \textbf{294}, 2566--8 (2001).

\bibitem{petersen+al_01}
Petersen, R.~S., Panzeri, S. \& Diamond, M.~E.
\newblock Population coding of stimulus location in rat somatosensory cortex.
\newblock \emph{Neuron} \textbf{32}, 503--14 (2001).

\bibitem{puchalla+al_05}
Puchalla, J.~L., Schneidman, E., Harris, R.~A. \& Berry, II, M.~J.
\newblock Redundancy in the population code of the retina.
\newblock \emph{Neuron} \textbf{46}, 493--504 (2005).

\bibitem{narayanan+al_05}
Narayanan, N.~S., Kimchi, E.~Y. \& Laubach, M.
\newblock Redundancy and synergy of neuronal ensembles in motor cortex. 
\newblock \emph{J Neurosci} \textbf{25}, 4207--16 (2005).


\bibitem{schneidman+al_06}
Schneidman, E., Berry, II, M.~J., Segev, R. \& Bialek, W.
\newblock Weak pairwise correlations imply strongly correlated network states
  in a neural population.
\newblock \emph{Nature} \textbf{440}, 1007--12 (2006).

\bibitem{shlens+al_06}
Shlens, J. \emph{et~al.}
\newblock The structure of multi-neuron firing patterns in primate retina. 
\newblock \emph{J Neurosci} \textbf{26}, 8254--66 (2006).


\bibitem{chechik+al_06}
Chechik, G. \emph{et~al.}
\newblock Reduction of information redundancy in the ascending auditory
  pathway.
\newblock \emph{Neuron} \textbf{51}, 359--68 (2006).

\bibitem{osborne+al_08}
Osborne, L.~C., Palmer, S.~E., Lisberger, S.~G. \& Bialek, W.
\newblock The neural basis for combinatorial coding in a cortical population
  response. 
\newblock \emph{J  Neurosci} \textbf{28}, 13522--31 (2008).

\bibitem{shlens+al_09}
Shlens, J. \emph{et~al.}
\newblock The structure of large-scale synchronized firing in primate retina. 
\newblock \emph{J Neurosci} \textbf{29}, 5022--31 (2009).

\bibitem{tkacik+al_09}
Tka$\check{\textrm{c}}$ik, G., Schneidman, E., Berry, II, M.~J. \& Bialek, W.
\newblock Spin glass models for a network of real neurons. 
arXiv:0912.5409 (2009).

\bibitem{schneidman+al_11}
Schneidman, E. \emph{et~al.}
\newblock Synergy from silence in a combinatorial neural code.
\newblock \emph{J  Neurosci} \textbf{31}, 15732--41 (2011). 


\bibitem{soo+al_11}
Soo, F.~S., Schwartz, G.~W., Sadeghi, K. \& Berry, II, M.~J.
\newblock Fine spatial information represented in a population of retinal
  ganglion cells.
\newblock \emph{J Neurosci} \textbf{31}, 2145--55 (2011). 


\bibitem{doi+al_12}
Doi, E. \emph{et~al.}
\newblock Efficient coding of spatial information in the primate retina. 
\newblock \emph{J Neurosci} \textbf{32}, 16256--64 (2012).

\bibitem{tishby+al_99}
Tishby, N., Pereira, F.~C. \& Bialek, W.
\newblock The information bottleneck method.
\newblock \emph{Proceedings of the 37th Annual Allerton Conference on
  Communication, Control and Computing,} \textbf{37}, 368--377 (1999).

\bibitem{creutzig+al_09}
Creutzig, F., Globerson, A. \& Tishby, N.
\newblock Past-future information bottleneck in dynamical systems.
\newblock \emph{Phys Rev E} \textbf{79}, 041925 (2009). 

\bibitem{barlow_53}
Barlow, H.~B.
\newblock Summation and inhibition in the frog's retina.
\newblock \emph{J Physiol (Lond)} \textbf{119}, 69--88 (1953). 

\bibitem{lettvin+al_60}
Lettvin, J.~Y., Maturana, H.~R., McCulloch, W.~S. \& Pitts, W.~H.
What the frog's eye tells the frog's brain.  {\emph Proc IRE} {\bf 47}, 1940--51 (1959).

\bibitem{maturana+al_60}
Maturana, H.~R., Lettvin, J.~Y., McCulloch, W.~S. \& Pitts, W.~H.
\newblock Anatomy and physiology of vision in the frog ({\em Rana pipiens}).
\newblock \emph{J Gen Physiol} \textbf{43}, 129--75 (1960). 

\bibitem{barlow+al_64}
Barlow, H.~B., Hill, R.~M. \& Levick, W.~R.
\newblock Retinal ganglion cells responding selectively to direction and speed
  of image motion in the rabbit.
\newblock \emph{J Physiol (Lond)} \textbf{173}, 377--407 (1964). 

\bibitem{roth_book}
Roth, G.
\newblock \emph{Visual Behavior in Salamanders} 
(Springer--Verlag, Berlin, 1987).

\bibitem{lee+al_93}
Lee, B.~B., Wehrhahn, C., Westheimer, G. \& Kremers, J.
\newblock Macaque ganglion cell responses to stimuli that elicit hyperacuity in
  man: detection of small displacements.
\newblock \emph{J  Neurosci} \textbf{13}, 1001--9 (1993). 

\bibitem{frechette+al_05}
Frechette, E.~S. \emph{et~al.}
\newblock Fidelity of the ensemble code for visual motion in primate retina.
\newblock \emph{J Neurophysiol} \textbf{94}, 119--35 (2005). 

\bibitem{thiel+al_07}
Thiel, A., Greschner, M., Eurich, C.~W., Ammermuller, J. \& Kretzberg, J.
\newblock Contribution of individual retinal ganglion cell responses to
  velocity and acceleration encoding.
\newblock \emph{J  Neurophysiol} \textbf{98}, 2285--96 (2007). 

\bibitem{bialek+al_06}
Bialek, W., de~Ruyter~van Steveninck, R.~R. \& Tishby, N.
\newblock Efficient representation as a design principles for neural coding and
  computation.
\newblock \emph{Proceedings of the International Symposium on Information
  Theory}   (2006).

\bibitem{brenner+al_00}
Brenner, N., Strong, S.~P., Koberle, R., Bialek, W. \& de~Ruyter~van
  Steveninck, R.~R.
\newblock Synergy in a neural code.
\newblock \emph{Neural comp} \textbf{12}, 1531--52 (2000).

\bibitem{mackay+mcculloch_52}
MacKay, D. \& McCulloch, W.
\newblock The limiting information capacity of a neuronal link.
\newblock \emph{Bull Math Biophys} \textbf{14}, 127--35 (1952).

\bibitem{rieke+al_93}
Rieke, F., Warland, D. \& Bialek, W.
\newblock Coding efficiency and information rates in sensory neurons.
\newblock \emph{Europhys Lett} \textbf{22}, 151--6 (1993). 

\bibitem{barlow_61}
Barlow, H.~B.
\newblock Possible principles underlying the transformation of sensory
  messages. In
\newblock \emph{Sensory Communication}, W Rosenblith, ed, pp 217--234 (Wiley, New York, 1961).

\bibitem{attneave_54}
Attneave, F.
\newblock Some informational aspects of visual perception.
\newblock \emph{Psych Rev} \textbf{61}, 183--93 (1954). 

\bibitem{atick+redlich_92}
Atick, J.~J. \& Redlich, A.~N.
\newblock What does the retina know about natural scenes?
\newblock \emph{Neural Comp} \textbf{4}, 196--210 (1992).

\bibitem{dan+al_96}
Dan, Y., Atick, J.~J. \& Reid, R.~C.
\newblock Efficient coding of natural scenes in the lateral geniculate nucleus:
  Experimental test of a computational theory.
\newblock \emph{J Neurosci} \textbf{16}, 3351--62 (1996). 

\bibitem{srinivasan+al_82}
Srinivasan, M.~V., Laughlin, S.~B. \& Dubs, A.
\newblock Predictive coding: a fresh view of inhibition in the retina.
\newblock \emph{Proc  R Soc Lond Ser B}
  \textbf{216}, 427--59 (1982). 
  
\bibitem{berry+schwartz_11}
Berry, II, M.~J. \& Schwartz, G.
\newblock The retina as embodying predictions about the visual world.
\newblock \emph{Predictions in the Brain:\ Using Our Past to Generate a Future}, M Bar, ed, pp
  295--310 (Oxford University Press, 2011).

\bibitem{berry+al_99}
Berry, II, M.~J., Brivanlou, I.~H., Jordan, T.~A. \& Meister, M.
\newblock Anticipation of moving stimuli by the retina.
\newblock \emph{Nature} \textbf{398}, 334--8 (1999).

\bibitem{schwartz+al_07}
Schwartz, G., Harris, R., Shrom, D. \& Berry, II, M.~J.
\newblock Detection and prediction of periodic patterns by the retina.
\newblock \emph{Nature Neurosci} \textbf{10}, 552--4 (2007). 

\bibitem{strong+al_98}
Strong, S.~P., Koberle, R., van Steveninck, R. R.~D. \& Bialek, W.
\newblock Entropy and information in neural spike trains.
\newblock \emph{Phys Rev Lett} \textbf{80}, 197--200 (1998). 

\bibitem{chechik+al_05}
Chechik, G., Globerson, A., Tishby, N. \& Weiss, Y.
\newblock Information bottleneck for gaussian variables.
\newblock \emph{JMLR} \textbf{6}, 165--88 (2005).

\end{thebibliography}
\end{document}